\documentclass[10pt,aps,ams,nofootinbib,showpacs,preprintnumbers]{revtex4}

\usepackage{enumitem}
\usepackage{graphicx}
\usepackage{amsmath}
\usepackage{amsfonts}
\newcommand{\be}{\begin{equation}}
\newcommand{\ee}{\end{equation}}
\newcommand{\bea}{\begin{eqnarray}}
\newcommand{\eea}{\end{eqnarray}}
\newcommand{\bean}{\begin{eqnarray*}}
\newcommand{\eean}{\end{eqnarray*}}
\usepackage{latexsym}
\usepackage[T1]{fontenc}

\begin{document}
\title{Relational evolution of a simple quantum Hamiltonian model} 

\author{Daniele Colosi}
\email[Email address: ]{colosi@matmor.unam.mx}
\affiliation{Instituto de Matem\'aticas, UNAM, Campus Morelia\\ C.P. 58190, Morelia, Michoac\'an, Mexico}
\date{\today}

\begin{abstract}
We study the quantum dynamics of a time reparametrization invariant system with a vanishing Hamiltonian. The evolution of the physical degrees of freedom of the system is described, both at the classical and at the quantum level, in relational terms by the construction of an internal time parameter. We use the Pegg-Barnett phase operator formalism in finite dimensional Hilbert space as an essential ingredient. 
\end{abstract}

\maketitle

\section{Introduction}
The description and the physical interpretation of the quantum dynamics of general covariant systems appear problematic due to the absence of a clear and unambiguous notion of time for such systems. This lack of a single preferred independent external time variable contrasts with the standard formulation of quantum mechanics where time plays the r\^ole of an absolute element, described by a classical parameter and not by an operator and is required for the description of the quantum evolution (it explicitly appears in the Schr\"odinger equation)\footnote{In the context of field theory, the formulation of the quantum theory uses not only a classical time variable but a whole fixed, non-dynamical background spacetime metric, the Minkowski metric.}. This tension between generally covariant systems and standard quantum mechanics poses both technical and conceptual problems. This kind of difficulties are particularly evident when one tries to quantize general relativity. The background independence of the gravitational field, consequence of the fundamental symmetry of the theory, namely general covariance, represents a major obstacle to the application of the standard background dependent methods of quantization of classical fields. However techniques have been developed to preserve background independence at the quantum level: Loop Quantum Gravity, a candidate for the theory of quantum gravity, realizes a (canonical) quantization of the gravitational field in a background independent way \cite{book}. Concerning the conceptual issues, several solutions have been proposed in order to provide a consistent physical interpretation of quantum generally covariant systems. In this letter we want to study one of these proposals advanced by Rovelli \cite{Rovelli90,Rovelli91,Rovelli91-2,book}, consisting in describing evolution of the physical degrees of freedom of the quantum system in a relational manner\footnote{This proposal is also known as "evolving constants of motion".}. We implement such a relational description in a simple Hamiltonian model that shares with generally covariant system the property of being invariant under time reparametrization.

The paper is organized as follows. In the section II we introduce the model of two harmonic oscillators with an Hamiltonian constraint and study their properties. In particular we provide a definition of an internal time that allows us to describe the classical evolution in a relational manner. In section III we shortly present the canonical quantization of the model and its physical Hilbert space (all the details can be found in the references cited). In section IV we translate at the quantum level the classical relational dynamics presented in section II. A fundamental ingredient will be the Pegg-Barnett formalism that gives a consistent and well-defined exponential phase operator acting on finite dimensional Hilbert space. By means of this exponential phase operator we construct well-defined gauge invariant operators that represent physical observables of the quantum system and we implement at the quantum level the relational evolution of section II. As a by-product of the study of the spectrum of these gauge invariant operators we obtain some \textit{new} relations between the zeros of the Hermite polynomials.

\section{The model}
With the aim of study the problem of time in quantum general covariant systems, and in particular in quantum gravity, Rovelli presented in \cite{Rovelli90} a simple system consisting of two harmonic oscillators with a total fixed energy\footnote{We will follow the treatment presented in \cite{Montesinos}. See also \cite{pendolo} for an alternative description.}. The system has one physical degree of freedom and one first-class constraint \cite{Dirac}. The extended phase space $\Gamma$ is coordinatized by the canonical pairs $(q_i,p_i); i=1,2$. The symplectic two-form is $\omega = dp_i \wedge dq_i$. The pair $(\Gamma, \omega)$ forms a symplectic space. The extended configuration space is defined by the coordinates $q_i$, also called partial observables to distinguish them from the complete observables represented by gauge invariant correlations between partial observables. We refer to \cite{book} for details. The dynamics, both classical and quantum, may be described using a parameter $\tau$ which is not the physical time, it is not connected with observability but instead it plays the same role of the time coordinate in general relativity: physically meaningful quantities are independent from the time coordinate. The action of this system has the form
\bea
S[q_i,p_i,\lambda] = \int d\tau \left\{ \frac{dq_i}{d\tau} \, p_i - \lambda H(q_i,p_i) \right\},
\eea
where $\lambda$ is a Lagrangian multiplier. This action is invariant under reparametrizations of the parameter $\tau$. The equations of motion are obtained by the variation of the action with respect to the canonical coordinates $(q_i,p_i)$,
\bea
\frac{dq_i}{d \tau} = \lambda \frac{\partial H(q_i,p_i)}{\partial p_i}, \\
\frac{dp_i}{d \tau} = - \lambda \frac{\partial H(q_i,p_i)}{\partial q_i}, 
\eea
while the variation of the action with respect to the Lagrangian multiplier $\lambda$ gives the constraint equation
\bea
H(q_i,p_i)=0, \  \ i=1,2.
\label{c}
\eea
The constraint has the following form 
\bea
H(q_i,p_i) = H_1 + H_2 -M =0,
\label{1}
\eea
where $H_1=\frac{1}{2}p_1^2+\frac{1}{2}q_1^2$ is the Hamiltonian of the oscillator $1$, and respectively $H_2$ for the oscillator $2$, and $M$ is a constant.
The dynamics is then described by the vanishing Hamiltonian (\ref{1}) and therefore the system shares with generally covariant system like general relativity the important property of being invariant under time reparametrizations. The physical meaning of such property is that the time parameter appearing in the solution of the equation of motion (given below, equations (\ref{qi},\ref{pi})), namely the parameter $\tau$, does not represent a physical time. It is a pure gauge parameter. Therefore it is the unfolding of the gauge transformation generated by the first class constraint (\ref{1}) that defines the evolution of the degrees of freedom of the system, i.e. dynamics is gauge.
We can parametrize the solution of the equation of motion as
\bea
q_i(\tau) = A_i \sin(\tau+ \phi_i),\  \ i=1,2,
\label{qi}
\eea
and the momenta are given by
\bea
p_i(\tau) = A_i \sin(\tau+ \phi_i), \  \ i=1,2.
\label{pi}
\eea
These solutions describe an ellipse of radii $A_1$ and $A_2$ and inclination $\Delta \phi= \phi_1-\phi_2$ in the extended configuration space, the plane $(q_1,q_2)$, with $(A_1,A_2) \in \left[ 0, \sqrt{2M}\right]$ and $(\phi_1,\phi_2) \in \left[0, 2 \pi \right]$. The values of the constants $A_1,A_2$ are fixed by the constraint (\ref{c}) to satisfy $A_1^2+A_2^2=2M$. The physical predictions of this system regard the relation between $q_1$ and $q_2$, not the dependence of $q_1$ and $q_2$ on the unphysical parameter $\tau$. Consequently the partial observables $q_1$ and $q_2$ are not physical observables. Instead the physical observables fo the system are given by the so called complete observables, namely gauge invariant correlations between partial observables.

The Hamiltonian constraint (\ref{c}) defines a constraint surface in the extended phase space $\Gamma$. The restriction of the symplectic two-form $\omega = dp_i \wedge dq_i$ to such constraint surface is degenerate. Indeed the Hamiltonian vector field $X$ generated by the constraint and given by
\bea
X= p_i \frac{\partial}{\partial q_i} - q_i \frac{\partial}{\partial p_i},
\eea
satisfies $\omega(X)=0$. The integral lines of $X$ on the constraint surface define the "orbits" of $\omega$, namely the motions.
The constraint surface can be parametrized by the set of independent coordinates $(\tilde{q},\tilde{p}_,t)$, where $t$ coordinatize the orbits and $(\tilde{q},\tilde{p})$ are canonical variables that coordinatize the physical phase space $\gamma$. Of course they satisfy $\{ \tilde{q}, \tilde{p} \} =1$ on $\gamma$, and they represent physical observables. The symplectic form on $\gamma$ is $\omega = d \tilde{p} \wedge d \tilde{q}$. Therefore the solution of the equation of motion (\ref{qi},\ref{pi}) can be expressed as
\bea
q_i=q_i(\tilde{q}, \tilde{p}; t), \label{qq} \\
p_i=p_i(\tilde{q}, \tilde{p}; t).
\label{pp}
\eea
On the other hand, the canonical variables $\tilde{q}$ and $\tilde{p}$ are functions of the coordinates of the extended phase space, 
\bea
\tilde{q}= \tilde{q}(q_i,p_i), \label{qq1} \\
\tilde{p}= \tilde{p}(q_i,p_i),
\label{pp1}
\eea
where the explicit dependence is fixed by the solution of the equation of motion. The orbit coordinate $t$ as well can be expressed as a function of $q_i$ and $p_i$:
\bea
t=t(q_i,p_i).
\label{t}
\eea
This parameter can be interpreted as an internal time with respect to which we describe the evolution of the system. The number of such internal time variables is fixed by the constraint: In our case there is only one constraint, so we have one parameter $t$. An interesting property of the internal time variable is that the expression (\ref{t}) is in general a multi-valued function, in contrast to the monotonically increasing time variable of non generally covariant systems. As a consequence, the values of the variable $t$ can lead to an interpretation of the evolution of the degrees of freedom of the system as forward or backward in time.
Finally, the relational description of the dynamics of the system is encoded in the combination of the solution of the equations of motion (\ref{qq}) and (\ref{pp}) with the definition of the internal time variable (\ref{t}).

We will choose the following canonical variables on the physical phase space
\bea
\tilde{q}(q_i,p_i) &=& \frac{1}{2}(q_1^2 + p_1^2) = A_1^2, \label{qtilde} \\
\tilde{p}(q_i,p_i) &=& \arctan \frac{q_1}{p_1} - \arctan \frac{q_2}{p_2} = \Delta \phi. \label{ptilde}
\eea
It is easy to show that $A_1^2$ and $\Delta \phi$ are conjugate variables, namely $\{A_1^2, \Delta \phi \} = 1$. They represent physical quantities: They are indeed independent of the gauge parameter $\tau$ and have vanishing Poisson brackets with the constraint (\ref{c}). The variable $t$ is given by
\bea
t = \arctan \frac{q_2}{p_2}.
\label{t1}
\eea
As already said this variable plays the r\^ole of an internal time for the system: It is the parameter with respect to which we describe the evolution of the degree of freedom of the system in a gauge invariant way. We notice that the time variable (\ref{t1}) is not a physical observale, indeed its Poisson bracket with the Hamiltonian constraint does not vanish. In particular it results to be equal to 1: $\{t,H\}=1.$ Furthermore, the internal time variable (\ref{t1}) is a multivalued function of $q_2,p_2$ because the arctangent only defines $t$ mod $2 \pi$.
Of course this particular choice of coordinatizing the constraint surface is not at all mandatory. We can perfectly choose different variables, for example the couple $(A_2^2, \Delta \phi)$ with $t = \arctan \frac{q_1}{p_1}$. 

If we plug the choice of variables (\ref{qtilde},\ref{ptilde}) and the internal time (\ref{t1}) into the solution (\ref{qq},\ref{pp}), we obtain
\bea
q_1(A_1^2, \Delta \phi,t) &=& \sqrt{A_1^2} \, \sin(t + \Delta \phi), \label{q1}\\
q_2(A_1^2, \Delta \phi,t) &=& \sqrt{M - A_1^2} \, \sin(t - \Delta \phi),
\eea
and the momenta
\bea
p_1(A_1^2, \Delta \phi,t) &=& \sqrt{A_1^2} \, \cos(t + \Delta \phi), \\
p_2(A_1^2, \Delta \phi,t) &=& \sqrt{M - A_1^2} \, \cos(t - \Delta \phi). \label{p2}
\eea
The set of equations (\ref{q1}-\ref{p2}) give the relational evolution of the $q_i$ and $p_i$ with respect to the internal time (\ref{t1}) for the particular choice of physical observales $(A_1^2, \Delta \phi)$. Indeed, choosen a particular combination of a set of variables of the extended phase space $\Gamma$ as an internal time parameter (in our case the couple $(q_2,p_2)$ defining the internal time (\ref{t1})), we can describe the evolution of the remaining set of variables, namely $q_1$ and $p_1$, as functions of this combination (\ref{t1}) of $q_2,p_2$ for any physical state $(A_1^2, \Delta \phi)$ of the system. A different interpretation is viable. The expressions (\ref{q1}-\ref{p2}) give a one-parameter family of physical observables defined on the physical phase space, where the parameter is the particular combination of $q_2,p_2$ given by (\ref{t1}). We can then say that (\ref{q1}-\ref{p2}) are evolving constants of motion in the sense of Rovelli \cite{Rovelli90,Rovelli91,Rovelli91-2,book}.


\section{Quantum theory}

The canonical quantization of the system is straightforward: Upon quantization the constraint (\ref{1}) becomes a quantum Hamiltonian constraint $\hat{H}$, and the (classical) Hamiltonians of the two oscillators become positive definite operators $\hat{H}_1$ and $\hat{H}_2$. The physical Hilbert space ${\cal H}_{phys}$ of the system is the space of states solutions of the Wheeler-DeWitt equation that defines the quantum dynamics: $\hat{H} | \Psi \rangle =0$. This Hilbert space turns out to be of finite dimension precisely because of the existence of the Hamiltonian constraint and the positive definiteness of $\hat{H}_1$ and $\hat{H}_2$. It is convenient to take the dimensionality of ${\cal H}_{phys}$ equal to the integer $2j+1= \frac{M}{\hbar}$. This finite dimensionality of the Hilbert space will be important in the following.

It is useful to introduce the creation and annihilation operators for the two oscillators\footnote{We use the letter $a$ and $b$ to indicate the annihilation operators for the oscillators 1 and 2 respectively. We take $\hbar=1$ in the rest of the paper.}, with the usual commutators $[a,a^{\dagger}]=[b,b^{\dagger}]=1$, $[a,b]=[a,b^{\dagger}]=0$, and the energy eigenstates $|n_1,n_2\rangle$ where $n_1$ and $n_2$ are the numbers of quanta of the oscillators $q_1$ and $q_2$ respectively, namely $n_1$ and $n_2$ are the eigenvalues of the number operators $\hat{N_1}=a^{\dagger} a$ and $\hat{N_2}=b^{\dagger}b$ respectively. The quantum Hamiltonian constraint $\hat{H}$ can be expressed in terms of these operators,
\bea
\hat{H} = \hat{N_1}+ \hat{N_2} +1 - M.
\label{qhc}
\eea
Therefore the physical Hilbert space is spanned by the vectors satisfying the equation $M=n_1+n_2+1$. It is convenient to introduce the quantum number $m = \frac{1}{2}(n_1-n_2)$ that runs from $m=-j$ to $m=j$, with $j=(M-1)/2$. So that ${\cal H}_{phys}$ is spanned by the $(2j+1)$ states $|m\rangle \equiv |j-m,j+m \rangle$. The representation of the states in the space of coordinates is
\bea
\psi_m(q_1,q_2)= \left(2^{2j}\pi (j+m)!(j-m)!\right)^{-1/2} H_{j+m}(q_1)H_{j-m}(q_2)e^{-\frac{q_1^2+q_2^2}{2}},
\eea
where $H_n(q)$ id the $n$-th Hermite polynomial. The dynamics can be obtained from an orthogonal projection operator $P$ from the kinematical Hilbert space $\cal K$ into the physical Hilbert space $\cal H$, $P: \cal K \rightarrow \cal H$, defined by
\bea
P= \int d\tau e^{-i \tau  H} = \sum_{m=-j}^j |m \rangle \langle m|,
\label{P}
\eea
and the propagator $K$ of the model is the integral kernel of $P$. The properties and the classical limit of the propagator $K$ have been studied in \cite{pendolo}. Moreover, following a proposal for computing the graviton propagator presented by Rovelli in \cite{Rovelli2005} and making use of the coherent states for this system derived in \cite{pendolo}, the two point function has been computed in \cite{2p}.

Here we adopt a different point of view. In order to describe the relational evolution at the quantum level, we want to construct the quantum version of the classical solutions (\ref{q1}-\ref{p2}). To do this we will take advantage of a property, mentioned above, of the physical Hilbert space, namely its finite dimensionality. This property results to be essential for the expression of the quantum operators corresponding to the canonical conjugate variables $(A_1^2, \Delta \phi)$ of the physical phase space of the theory introduced in the previous section. It turns out that such quantum operators can be defined in a consistent way.

\section{Relation evolution at the quantum level}
\subsection{Complete observables in the Pegg-Barnett formalism}
In a series of papers \cite{PB1,PB2,PB3} Pegg and Barnett constructed a well-defined Hermitian optical phase operator in a finite dimensional Hilbert space. Studying the quantum theory of harmonic oscillator in a finite dimensional Hilbert space, they succeeded in defining quantum phase states as well as a phase state operator.  We use the Pegg-Barnett phase formalism to write the Hermitian operator acting on the $(2j+1)$-dimensional ${\cal H}_{phys}$ corresponding to the physical observable $\Delta \phi$. Moreover this formalism enables us to construct a unitary exponential phase operator $\exp(\pm i \hat{\phi})$ (see references \cite{PB1,PB2,PB3} for the details of the formalism). Therefore we can consider the following operators,
\bea
\hat{q}_1(t)&=& \frac{1}{\sqrt{2}}\left(   e^{it} \,  (\widehat{N}_1)^{1/2}  \, e^{-i \widehat{\Delta \phi}}  + e^{-it} \, e^{i \widehat{\Delta \phi}} \, (\widehat{N}_1)^{1/2} \right), \label{q2hat}\\
\hat{p}_1(t)&=& \frac{i}{\sqrt{2}}\left( e^{it} \,  (\widehat{N}_1)^{1/2}  \, e^{-i \widehat{\Delta \phi}} -  e^{-it} \, e^{i \widehat{\Delta \phi}} \, (\widehat{N}_1)^{1/2}  \right), \label{p2hat}
\eea
where $\widehat{N}_1$ is the number operator of the quantum oscillator 1, namely $\widehat{N}_1=a^{\dagger}a$. The operator $e^{i \widehat{\Delta \phi}}$ has the following number state representation \cite{PB1,PB2,PB3}
\bea
e^{i \widehat{\Delta \phi}} = |-j \rangle \langle -j+1| +|-j+1 \rangle \langle -j+2| + \dots + |j-1 \rangle \langle j|+|j \rangle \langle -j| .
\label{ephi}
\eea
The parameter $t$ appearing in the above expressions is the same parameter defined in section II: The internal time given by (\ref{t1}).

The main property of the operators $\hat{q}_1(t)$ and $\hat{p}_1(t)$ is represented by their commutations relations with the quantum Hamiltonian constraint $\hat{H}$ (\ref{qhc}). With the expression (\ref{ephi}) of the exponential phase operator it is easy to show that both $\hat{q}_1(t)$ and $\hat{p}_1(t)$ commute with $\hat{H}$,
\bea
[ \hat{q}_1(t),\hat{H} ] = [\hat{p}_1(t),\hat{H} ] =0.
\eea
These operators are gauge invariant operators, representing therefore physical observables of the quantum system.

Introducing the operators $A=e^{i \widehat{\Delta \phi}} \, (\widehat{N}_1)^{1/2} $ and $A^{\dagger}=(\widehat{N}_1)^{1/2}  \, e^{-i \widehat{\Delta \phi}}$ the two above operators take the familiar form
\bea
\hat{q}_1(t) = \frac{1}{\sqrt{2}}\left( e^{it} \, A^{\dagger} +  e^{-it} A  \right), \  \ \hat{p}_1(t)= \frac{i}{\sqrt{2}}\left( e^{it} \, A^{\dagger} -  e^{-it} A  \right).
\eea
$A$ and $A^{\dagger}$ play the r\^ole of annihilation and creation operators acting on the finite dimensional physical Hilbert space. Their commutation relation is
\bea
[A,A^{\dagger}] = I - (2j+1) |j \rangle \langle j|,
\eea
where $I$ is the identity operator. The difference with the usual canonical commutation relations is a consequence of the finite dimensionality of the Hilbert space, and is at the heart of the Pegg-Barnett formalism. It is straightforward to calculate the commutation relation between the operators (\ref{q2hat},\ref{p2hat})
\bea
[\hat{q}_1(t),\hat{p}_1(t)] &=& i(A \, A^{\dagger} - A^{\dagger} \, A), \nonumber\\
&=& i (I - (2j+1) |j \rangle \langle j|).
\label{CRqp}
\eea
Once agian, the above commutator differs from the usual canonical commutator because of the finite dimensionality of the Hilbert space. We can write the equations of motion governing the evolution of these operators with respect to the internal time $t$ in Heisenberg form as
\bea
\frac{d \, \hat{q}_1(t)}{dt} &=& i [\hat{H}'_1,\hat{q}_1(t) ],\\
\frac{d \, \hat{p}_1(t)}{dt} &=&i [\hat{H}'_1,\hat{q}_1(t) ],
\eea
with $\hat{H}'_1 = A^{\dagger}A = \hat{N}_1$. This Hamiltonian differs from the Hamiltonian $\hat{H}_1$ of the oscillator 1 by a factor $1/2$. This difference has consequence in the classical limit of the propagator, see section IV.C. In the next section we will study the spectrum of these operators that turns out to be finite and discrete.

\subsection{Spectrum of the complete observables}
The Heisemberg equations of motion for the gauge invariant operators $\hat{q}_1(t)$ and $\hat{p}_1(t)$, with solution given by the expressions (\ref{q2hat}) and (\ref{p2hat}) respectively, represent the equations of motion of a one dimensional harmonic oscillator evolving with the time $t$. The generator of this time evolution is the Hamiltonian $\hat{H}'_1$. Its eigenvalues are the eigenvalues of the number operator of the oscillator 1, namely $(j+m)$. We can write the solution of the equations of motion adopting the Schr\"odinger picture: The kets solution of the quantum dynamics have following form
\bea
|q_k(t)\rangle = N_k \sum_{m=-j}^j (2^{j+m}(j+m)!)^{-1/2} H_{j+m}(q_k) e^{-q_k^2/2} e^{i(j+m)t} |m \rangle,
\label{q2t}
\eea
where $N_k$ is a normalization factor and $H_{j+m}$ is the Hermite polynomial of order $j+m$. In order to truncate the series at the value $m=j$, which corresponds to the maximum allowable eigenvalue of $\hat{N}_1$, the argument $q_k$ of the Hermite polynomial must be a zero of $H_{2j+1}$. There exist $2j+1$ of such zeros (labeled by the index $k$), consequently the states (\ref{q2t}) span the $(2j+1)$-dimensional physical Hilbert space. We recover the same result of \cite{Lawrie} and \cite{T1,T2}. The calculation the normalization factor $N_i$ is straightforward: 
\bea
\langle q_k(t) |q_k(t)\rangle  =|N_k|^2 \sum_{m=0}^{2j} \frac{1}{2^{m}m!} H_{m}^2(q_k)e^{-q_k^2},
\eea
therefore 
\bea
N_k= e^{q_k^2/2} \left( \sum_{m=0}^{2j} \frac{1}{2^{m}m!} H_{m}^2(q_k) \right)^{-1/2}. \label{Ni}
\eea

The $2j+1$ states $|q_k(t)\rangle, (k=1,...,2j+1)$, given by (\ref{q2t}) are characterized by different properties: 
\begin{enumerate}[label=(\roman{*})]
\item They are physical states of the quantum system. They are indeed solution of the Wheeler-DeWitt equation,
\bea
\hat{H}|q_k(t)\rangle=0,
\eea
with $\hat{H}$ given by (\ref{qhc}). In particular the states $|q_k(t)\rangle$ are linear combinations of the energy eigenstates $|m\rangle$. 
\item They also satisfy a Schr\"odinger equation describing their evolution with respect to the internal time parameter $t$ generated by the Hamiltonian $\hat{H}'_1$,
\bea
i \frac{\partial}{\partial t}|q_k(t)\rangle = \hat{H}'_1 |q_k(t)\rangle.
\label{Seq}
\eea
\item They are eigenstates of the operator $\hat{q}_1(t)$ defined by (\ref{q2hat}). Their $2j+1$ eigenvalues $q_k$ correspond to the zeros of the Hermite polynomial of order $2j+1$.
\item They undergo a cyclic evolution, in fact $|q_k(t+2 \pi)\rangle = |q_k(t)\rangle$.
\item They form an orthonormal basis of the physical Hilbert space, indeed 
\bea
\langle q_l(t) |q_k(t)\rangle =  N_k \, \overline{N_l} \,  \frac{H_{2j+1}(q_k)H_{2j}(q_l)-H_{2j}(q_k)H_{2j+1}(q_l)}{2^{2j}(2j)!(q_k-q_l)} \, e^{-(q_k^2+q_l^2)/2},
\eea
and $q_k$ and $q_l$ are zeros of the Hermite polynomial of order $2j+1$. Therefore the above expression is zero for $k \neq l$. Consequently we have 
\bea
\langle q_l(t) |q_k(t)\rangle = \delta_{kl}.
\eea 
Notice that the right side is the Kronecker $\delta$ and not the Dirac distribution. 
\item The transition amplitude from $|q_k(t)\rangle$ to $|q_l(t')\rangle$ is
\bea
\langle q_l(t') |q_k(t)\rangle=  N_k \, \overline{N_l} \,  \sum_{m=0}^{2j} \frac{1}{2^{m}m!} H_{m}(q_k)H_{m}(q_l) e^{-(q_k^2+q_l^2)/2} e^{im(t-t')}.
\label{prop}
\eea
This quantity may receive a clear physical interpretation. The modulus square of (\ref{prop}) represents the probability of measure the value $q_l$ of the physical observable $\hat{q}_1$ at time $t'$ if we have measure the value $q_k$ at time $t$.
\end{enumerate}

The property (v) allows to express the physical states in the basis of the energy eigenstates, $|m\rangle$, in terms of the physical states $|q_k(t)\rangle$,
\bea
|m\rangle &=& \sum_{k=1}^{2j+1} |q_k(t)\rangle \langle q_k(t) |m\rangle , \nonumber\\
&=& \sum_{k=1}^{2j+1} |q_k(t)\rangle \, \overline{N_k}  \, (2^{j+m}(j+m)!)^{-1/2} H_{j+m}(q_k) e^{-\frac{q_k^2}{2}} e^{-i(j+m)t}. \label{m-q}
\eea
Of course the state $|m\rangle$ is independent of the parameter $t$: The $t$-dependence of $|q_k(t)\rangle$ is exactly canceled by the last exponential in the above expression.
The scalar product between two physical states $|m \rangle$ and $|n \rangle$, expressed in the basis of the states $|q_k(t) \rangle$, leads to some \textit{new} relations between the zeros of the Hermite polynomials reported in the appendix.

\subsection{The classical limit}
In order to study the classical limit, we need the expression of the $2j+1$ eigenvalues $q_k$, roots of the Hermite polynomial $H_{2j+1}$, in the limit $j \rightarrow \infty$ (corresponding to the classical limit, see \cite{pendolo}).  Such expressions have been derived in the appendix of \cite{T1}: The roots the Hermite polynomial have the following asymptotic form
\bea
q_k &=& \frac{\pi k}{\sqrt{2(2j+1)+3}} \sqrt{1 + \frac{\pi^2 k^2 -3/2}{3(2(2j+1)+3)^2}} + O((2j+1)^{-4.5}), \ \ \hbox{for $2j+1$ odd,} \label{qinf} \\
q_k &=& \frac{\pi (k-1/2)}{\sqrt{2(2j+1)+3}} \sqrt{1 + \frac{\pi^2 (k-1/2)^2 -3/2}{3(2(2j+1)+3)^2}} + O((2j+1)^{-4.5}), \ \ \hbox{for $2j+1$ even.} \label{qinf2}
\eea
We can see that the eigenvalues $q_k$ span an interval centered in the origin of the real line of length $O(\sqrt{2j+1})$. The distance between two successive eigenvalues is $O(1/\sqrt{2j+1})$. When $j$ tends to infinity the set of eigenvalues $q_k$ becomes dense in the real axis. However, if we naively take the limit $j \rightarrow \infty$, at finite values of the label $k$, the asymptotic expressions (\ref{qinf}) and (\ref{qinf2}) vanish and consequently they fail to reproduce the correct spectrum of the operator $\hat{q}_1$. This issue has been analized in \cite{T1, T2} and the proposed solution consists in a scaling of the label $k$ of the form
\bea
k = \frac{\sqrt{2(2j+1)}}{\pi} \, q +b,
\label{scaling}
\eea
where the constants $q$ and $b$ are real. It has been shown in \cite{T2} that the value of $b$ has no influence on the result obtained in the continuum limit. Its r\^ole is to control the convergence of limit to the correct result\footnote{The scaling (\ref{scaling}) can be modified by the addition of one more parameter useful for the convergence of the classical limit. We refer to \cite{T2} for the discussion of this point.}. With the prescription (\ref{scaling}) any real number $q$ can now be obtained in the classical limit.

It is interesting to study the classical limit of the propagator defined in (\ref{prop}). Recalling the Mehler's formula
\bea
\sum_{n=0}^{\infty} H_n(x) H_n(y) \, \frac{z^n}{2^n n!} = (1-z^2)^{-1/2} \exp \left( \frac{2xyz - (x^2+y^2) z^2}{1-z^2} \right),
\eea
the evaluation of the limit $j \rightarrow \infty$ of the propagator (\ref{prop}) leads to the expression
\bea
\langle q_l(t') |q_k(t)\rangle&=&  N_k \, \overline{N_l} \, e^{-i(t-t')/2} \times \nonumber\\
&& \times \sqrt{\frac{i}{2 \sin(t-t')}} \exp \left[ \frac{i}{2 \sin(t-t')} \left( 2 q_k q_l-(q_k^2+q_l^2)\cos (t-t') \right) \right].
\label{propclass}
\eea
We recognize in the second line of (\ref{propclass}) the propagator of a single harmonic oscillator. The presence of the exponential factor $e^{-i(t-t')/2}$ is a direct consequence of the form of the Hamiltonian governing the evolution of the system, namely $\hat{H}'_1 = \hat{N}_1$. As noted previously this Hamiltonian is not equal to the Hamiltonian of the oscillator 1 because of the absence of a constant term $1/2$: $\hat{H}_1 = \hat{N}_1 + 1/2$. This difference is responsible for the appearing of the exponential $e^{-i(t-t')/2}$ in (\ref{propclass}).

\subsection{The two point function}
With the definition of time given above, we can now define a vacuum state associated with the minimum eigenvalue of the hamiltonian $\hat{H}'_1$. Of course, a different choice of the time parameter will lead to a different and inequivalent vacuum state. This situation reproduces the well know problem of the definition of a vacuum, and more generally of a particle concept, in quantum field theory in curved spacetime. In particular the state $|m=j\rangle$ corresponds to the lower eigenvalue of $\hat{H}'_1$. Having a definition of the vacuum state we can compute the two point function for the operator $\hat{q}_1$ defined by the values $t$ and $t'$ of the internal time:
\bea
\langle j | \hat{q}_1(t')\hat{q}_1(t) |j \rangle = j \, e^{-i2j (t'-t)}.
\label{2p}
\eea
We recover the same result (apart from a factor 2) obtained in \cite{2p}: In particular the calculation of the two point function in \cite{2p} has been carried out using the coherent states of the quantum system defined in \cite{pendolo}. So, taking the coherent state corresponding to the vacuum state $|j \rangle$ and picked on the two points of the classical trajectory (namely the ellipse) defined by the two values $t$ and $t'$ of the internal time parameter, the calculation of the two point function based on the procedure applied in \cite{2p} gives (\ref{2p}).

\section{Discussion}
As mentioned in section II the model of two harmonic oscillators with dynamics fixed by the hamiltonian constraint (\ref{c}) was first introduced in \cite{Rovelli90} in order to study the problem of time for a time reparametrization invariant system. In particular in \cite{Rovelli90} a gauge invariant physical observable describing the evolution of a coordinate in the extended configuration space, the plane $(q_1,q_2)$, say $q_1$, in terms of the coordinate, $q_2$, was explicitly constructed simply by eliminating the arbitrary and non-physical parameter $\tau$ in the solution of the classical equations of motion (\ref{qi},\ref{pi}), and interpreting $q_2$ as a real parameter. This defines a set of physical observables $q_1(t)$ interpreted as the value of $q_1$ when $q_2$ has the value $t$ (for every real $t$). At the quantum level an operator corresponding to $q_1(t)$ has been explicitly constructed and it turns out to be only approximately self-adjoint, in particular only when restricted to a subspace of the physical Hilbert space called in \cite{Rovelli90} the Schr\"odinger regime characterized roughly by states described by a wave function peaked around a classical trajectory. Mathematically the non self-adjointness comes from a term contained in the expression of $q_1(t)$ of the form
\bea
\sqrt{M+2 L_z-t^2},
\label{tt}
\eea
where the operator $L_z=\frac{1}{2}(b^{\dagger}b-a^{\dagger}a)$, and $a,b$ are the creation and annihilation operators of section III. For large value of $t$, (\ref{tt}) becomes immaginary. This implies that $q_1(t)$ has complex eigenvalues and that evolution can not be unitary outside the Schr\"odinger regime.

This problem is completely solved by our choice for the operator $\hat{q}_1(t)$ given by (\ref{q2hat}), which is a well-defined self-adjoint gauge invariant operator for each value of the internal time parameter $t$ defined in (\ref{t1}). Of course this internal time $t$ differs from the parameter appearing in (\ref{tt}). The operator $\hat{q}_1(t)$ commutes with the Hamiltonian constraint and represents therefore a physical observable of the quantum system. Its spectrum has been computed and results to be finite and discrete with all the eigenvalues are real. In particular these eigenvalues coincide with the zeros of the Hermite polynomial of order $2j+1$. These eigenvalues are physical predictions of the theory. 

Moreover, the operator $\hat{q}_1(t)$ realizes the quantum version of the classical quantity $q_1(A_1^2, \Delta \phi,t)$ defined in (\ref{q1}). We can conclude that the physical observable $\hat{q}_1(t)$ implements the relational description of the quantum dynamics of the system of two harmonic oscillator, i.e. expressing the evolution of one oscillator in terms of the other. A key ingredient has been the construction of an internal time parameter. Such parameter, given by (\ref{t1}), depends only on the phase space variables of the oscillator 2. It is not a monotonically increasing function as the usual time in quantum mechanics. Instead it is a multivalued function of the canonical classical variables $(q_2,p_2)$ and as a consequence the physical states $|q_k \rangle$ undergo a cyclic evolution. Having an internal time, a description of the quantum dynamics in terms of a Schr\"odinger equation satisfied by the physical quantum states of the system is then available, see equation (\ref{Seq}). We then recover the standard formulation of time-dependent quantum mechanics. However this has to be intended in a relational way. Indeed in usual quantum mechanics the phase space variables describing oscillator 2, namely $q_2$ and $p_2$, would have been promoted to the status of quantum operators. Such approach has been adopted in \cite{pendolo}. Here we have studied the system in a different way. The variables $q_2$ and $p_2$ are not quantized, they instead provide a definition of the internal time. The evolution of the remaining phase space variables $q_1$ and $p_1$, both at the classical and at the quantum level, is expressed with respect to such internal time.

Finally, we have obtain as a by-product of the study of the spectrum of the gauge invariant operator (\ref{q2hat}) some new relations between the zeros of the Hermite polynomials, see formulas (\ref{A1},\ref{A2},\ref{A3}).

\appendix
\section{Relations between the zeros of the Hermite polynomials}

We have seen that the eigenvectors of the operator $\hat{q}_1$ form an orthonormal basis of the $(2j+1)$-dimensional physical Hilbert space ${\cal H}_{phys}$, and the associated eigenvalues $q_k$ are the zeros of the Hermite polynomial of order $2j+1$. Using these two properties we derive in this appendix some \textit{new} relations between the zeros of the Hermite polynomials.
Starting from the orthonormality of the eigenstates of $\hat{q}_1$ we derived the expression of the energy eigenstates $|m\rangle$ in the basis of the physical states $|q_k(t)\rangle$ in formula (\ref{m-q}),
\bea
|m\rangle = \sum_{k=1}^{2j+1} |q_k(t)\rangle \, \overline{N_k}  \, (2^{j+m}(j+m)!)^{-1/2} H_{j+m}(q_k) e^{-\frac{q_k^2}{2}} e^{-i(j+m)t}.
\eea
As already noted, the $t$-dependence of $|q_k(t)\rangle$ is exactly canceled by the exponential term $e^{-i(j+m)t}$ in the above expression, so that the state $|m\rangle$ is independent of the parameter $t$.
The scalar product between two physical states $|m \rangle$ and $|n \rangle$, expressed in the basis of the states $|q_k(t) \rangle$, takes the following form,
\bea
\langle n|m\rangle = \sum_{k=1}^{2j+1} |N_k|^2 \, (2^{j+m}(j+m)!)^{-1/2} H_{j+m}(q_k) e^{-\frac{q_k^2}{2}} \, (2^{j+n}(j+n)!)^{-1/2} H_{j+n}(q_k) e^{-\frac{q_k^2}{2}}.
\label{mn}
\eea
The energy eigenstates $|m\rangle$ form an orthonormal basis of the physical Hilbert space. The scalar product (\ref{mn}) reduces to Kronecker delta: $\langle n|m\rangle = \delta{m,n}$. Substituting in (\ref{mn}) the expression of the normalization factor $N_k$ calculated in (\ref{Ni}) we arrive at
\bea
\delta_{nm}= \sum_{k=1}^{2j+1}(2^{n+m}\, m! \, n!)^{-1/2}H_{m}(q_k)H_{n}(q_k) \left( \sum_{l=0}^{2j} \frac{1}{2^{l}l!} H_{l}^2(q_k) \right)^{-1}.
\label{A1}
\eea
Let's consider this formula for some specific values of $m$ and $n$. Consider first the case $m=n=0$. We obtain
\bea
\sum_{k=1}^{2j+1} \frac{1}{ \sum_{l=0}^{2j} \frac{1}{2^{l}l!} H_{l}^2(q_k)}=1.
\label{A2}
\eea
For $m \neq 0$ and $n=0$,
\bea
\sum_{k=1}^{2j+1} \frac{1}{\sqrt{2^m \, m!}} \frac{H_{m}(q_k)}{ \sum_{l=0}^{2j} \frac{1}{2^{l}l!} H_{l}^2(q_k)}=1, \  \ m \in \{1,...,2j\}.
\label{A3}
\eea
The properties of the equations (\ref{A1},\ref{A2},\ref{A3}) will be studied elsewhere.

\end{document}